\documentclass[prd,showpacs,showkeys,superscriptaddress]{revtex4}

\usepackage{amssymb}

\usepackage{amsmath}
\usepackage[T1]{fontenc}
\usepackage[english]{babel}
\usepackage{graphicx}
\usepackage[latin9]{inputenc}
\usepackage{natbib}
\usepackage{rotating}

\begin{document}
\title{The generalized second law in universes with quantum corrected entropy relations}
 \author{Ninfa Radicella}
 \email{ninfa.radicella@uab.cat}
\affiliation{Departamento de F\'isica, Universidad Aut\'onoma de
Barcelona, 08193 Bellaterra (Barcelona), Spain} \affiliation{INFN,
Sezione di Torino, I-10100 Torino, Italy}
\author{Diego Pav\'on}
\email{diego.pavon@uab.es} \affiliation{Departamento de F\'isica,
Universidad Aut\'onoma de Barcelona, 08193 Bellaterra (Barcelona),
Spain}

\begin{abstract}
We apply the generalized second law of thermodynamics to
discriminate among quantum corrections (whether logarithmic or
power-law) to the entropy of the apparent horizon in spatially
Friedmann-Robertson-Walker universes. We use the corresponding
modified Friedmann equations along with either Clausius relation
or the principle of equipartition of the energy to set limits on
the value of a characteristic parameter entering the said
corrections.
 \end{abstract}
\maketitle
\section{Introduction}
As is well known, event horizons, whether  black holes' or
cosmologicals, mimic black bodies and possess a nonvanishing
temperature and entropy, the latter obeying the Bekenstein-Hawking
formula, \cite{bekenstein73,davies75,gibbons77}
\begin{equation}\label{entropy}
S_{BH}=k_B\frac{A}{4l^2_{pl}} \, .
\end{equation}
\noindent This expression, in which $k_B$ stands for the Boltzmann
constant, $A$ the area of the horizon, and
$l_{pl}=\sqrt{G\hbar/c^3}$ the Planck's length, points to a deep
connection between gravitation, thermodynamics, and quantum
mechanics, still far from being fully unveiled though progress in
that direction are being made -see, e.g., \cite{wu08a,wu08b} and
references therein. Recently it was demonstrated that cosmological
apparent horizons are also endowed with thermodynamical
properties, formally identical to those of event horizons
\cite{cai09}.

The connection between gravity and thermodynamics was reinforced
by Jacobson, who associated Einstein equations with Clausius'
relation \cite{jacobson95}, and later on by Padmanabhan who linked
the macroscopic description of spacetime, given by Einstein
equations, to microscopic degrees of freedom, $N$, through the
principle of equipartition of energy, i.e.,
\begin{equation}\label{equipartition}
E=\frac{1}{2}N k_B T \, .
\end{equation}
In particular, Padmanabhan, starting from the field equations
arrived to the equipartition law \cite{padmanabhan10b} and
indicated how to obtain the field equations of any diffeormorphism
invariant theory of gravity from an entropy extremising principle
\cite{padmanabhan07}, the entropy
of spacetime being proportional to $N$. \\

On the other hand, quantum corrections to the semi-classical
entropy-law (\ref{entropy}) have been introduced in recent years,
namely, logarithmic and power-law corrections. Logarithmic
corrections, arises from loop quantum gravity due to thermal
equilibrium fluctuations and quantum fluctuations
\cite{meissner04,  ghosh04, chatterjee04},
$$
S\propto\left[\frac{A}{4l^2_{pl}}+\alpha
\ln{\frac{A}{4l^2_{pl}}}\right] \, .
$$
\noindent On its part, power-law corrections  appear in dealing
with the entanglement of quantum fields in and out the horizon
\cite{saurya08},
$$
S\propto\frac{A}{4l^2_{pl}}\left[1-K_\alpha A^{1-\alpha/2}\right]
\, .
$$
In the last two expressions, $\alpha$ denotes a dimensionless
parameter whose value (in both cases) is currently under debate.\\

For the connection between horizons and thermodynamics  to hold,
these quantum entropy corrections must translate into
modifications of the field equations of gravity, see e.g.
\cite{lidsey09,cai09,zhang10}. In any sensible cosmological context
these modifications must fulfill the generalized second law (GSL)
of thermodynamics. The latter asserts that the entropy of the
horizon plus the entropy of its surroundings must not decrease in
time. As demonstrated by Bekenstein, this law is satisfied by
black holes in contact with their radiation \cite{bekenstein74}.

The aim of this paper is to see whether the modified Friedmann
equations coming from logarithmic corrections and from power-law
corrections, in conjunction with Clausisus relation or the
equipartition principle, are compatible with the generalized
second law. This will set constraints on the parameter $\alpha$
introduced above, whose value is theory dependent and rather
uncertain. We hope this may be of help in discriminating among
quantum corrections, via a purely classical analysis.\\

The plan of this work is as follows: Section \ref{Modified Friedmann equations} derives Friedmann equations from different entropy
corrections, considering a Friedmann-Robertson-Walker (FRW) metric
sourced by a perfect fluid. Section \ref{Generalized second law of
Thermodynamics} considers the entropy rate enclosed by the
apparent horizon and in which the fluid is in thermal equilibrium
with it. Section \ref{GSL with phantom fluid or particle
production} e xtends the analysis by allowing the fluid to be
phantom, for a period, or by allowing the production of particles.
Finally, Section \ref{Conclusions} summarizes and discusses our
findings.

\section{Modified Friedmann equations}\label{Modified Friedmann equations}
In this section we recall the derivation of the modified Friedmann
equations in case of the classes of entropy corrections to the
Bekenstein-Hawking entopy in eq.(\ref{entropy}).
As said above, this semiclassical relation gets modified when quantum corrections are taken into account.\\
Logarithmic corrections lead to the expansion \cite{meissner04,
ghosh04, chatterjee04}
\begin{equation}\label{log}
S=k_B \, \left[\frac{A}{4l^2_{pl}}+\alpha
\ln{\frac{A}{4l^2_{pl}}}\right],
\end{equation}
while power-law corrections yield \cite{saurya08}
\begin{equation}\label{power}
S= k_B \, \frac{A}{4 l^2_{pl}}\left[1-K_\alpha
A^{1-\alpha/2}\right],
\end{equation}
for the horizon entropy. In (\ref{power}) $K_\alpha$ is a parameter that depends on the
power $\alpha$ of the entropy correction.\\
We wish to examine thermodynamical behavior of the system consisting in the apparent horizon of a spatially
flat FRW universe and the fluid within it.\\
The FRW metric can be written as
$$
ds^2=h_{ab}dx^a dx^b+\tilde{r}^2d\Omega^2,
$$
where $\tilde{r} = a(t) r$ and  $h_{ab} = \text{diag}(-1,a(t)^2)$.
The apparent horizon is defined by the condition $h^{ab}\partial_a
\tilde{r}\partial_b\tilde{r}=0$ so that its radius turns out to be
\begin{equation}\label{apphor}
\tilde{r}_{AH}=\frac{c}{H},
\end{equation}
where $H=\dot{a}/a$ denotes the Hubble function. From expressions
(\ref{log}) and (\ref{power}) different cosmological scenarios can
be considered, depending on whether use is made of Clausius
relation  \cite{jacobson95}
\begin{equation}
\delta Q= TdS  \, ,
\label{clausius}
\end{equation}
or the principle of equipartition of the energy, $d E=k_B d( N
T)/2$ \cite{padmanabhan10a}.  In either case, the left hand side
represents the amount of energy that crosses the apparent horizon
within a time interval $dt$ in which the apparent horizon evolves
from $\tilde{r}_{AH}$ to $\tilde{r}_{AH}+d \tilde{r}_{AH}$
$$
dE= A_{AH} T_{\mu\nu}k^\mu k^\nu dt \, .
$$
Here $T_{\mu\nu}=(\rho+P/c^2) u_\mu u_\nu +  P g_{\mu\nu}/c^2$ is
the energy-momentum tensor of the perfect fluid, and $k^a$ is the
approximate generator of the horizon, $k^a=(1,-Hr,0,0)$. It
follows that
\begin{equation}\label{energy}
dE=4\pi \tilde{r}^3\left(\rho+\frac{P}{c^2}\right)\ H\ dt.
\end{equation}
The change of the area of the apparent horizon, $dA_{AH}=8\pi
\tilde{r}_{AH} d\tilde{r}_{AH}$, induces  the entropy shift,
\begin{equation}\label{shift}
dS=\frac{\partial S}{\partial A}dA=-\frac{8\pi
\tilde{r}_{AH}^4}{c^2}\frac{\partial S}{\partial A}H\dot{H} dt\, .
\end{equation}
The number of degrees of freedom is assumed proportional to the
entropy whereby it also changes, and the same holds true for the
temperature of the system, that we take as the temperature of the
horizon \cite{cai09}
\begin{equation}\label{temperature}
T_{H}=\frac{\hbar}{k_B}\frac{c}{2\pi\  \tilde{r}_{AH}}\, .
\end{equation}
By using eq.(\ref{energy}) with either Clausius relation
(\ref{clausius}) or the equipartition principle
(\ref{equipartition}), one gets the modified Friedmann equations,
\begin{eqnarray}
H^2\left[1+g(\alpha, H)\right]&=&\frac{8\pi G}{3} \rho,\label{H}\\
\dot{H}\left[1+f(\alpha, H)\right]&=&-4\pi G\left(\rho+\frac{P}{c^2}\right)\label{Hdot},
\end{eqnarray}
the explicit expressions of $f(\alpha, H)$ and $g(\alpha, H)$
depend on both the entropy corrections and the thermodynamical
relation employed, as shown in Table \ref{table1}. It should be
noted that, eq.(\ref{H}) can be recovered from eq.(\ref{Hdot}) and
the evolution equation for the perfect fluid:
\begin{equation}\label{conservation}
\dot{\rho}+3H\left( \rho+\frac{P}{c^2}\right)=0.
\end{equation}

\begin{table}
\caption{Expressions for $f(\alpha,H)$ and $g(\alpha,H)$.}
\begin{tabular}{|c||c||}
\hline
                        &Logarithmic correction\\
\hline\hline
Equipartition&$f(\alpha, H)=\frac{l^2_p\alpha}{2\pi c^2} H^2\left\{1-\frac{1}{2}\ln{\left(\frac{\pi c^2}{l^2_p H^2}\right)}\right\}$\\
                       &$g(\alpha, H)=\frac{3l^2_p\alpha}{16 \pi c^2} H^2\left\{1+\frac{2}{3}\ln{\left(\frac{\pi c^2}{l^2_p H^2}\right)}\right\}$\\
\hline
Clausius       &$f(\alpha, H)=\frac{l^2_p\alpha}{2\pi c^2} H^2$\\
                       &$g(\alpha, H)=\frac{3l^2_p\alpha}{16 \pi c^2} H^2$\\
\hline\hline
                       &Power-law correction\\
\hline\hline
Equipartition&$f(\alpha, H)=-\alpha\frac{3-\alpha}{4-\alpha} (Hr_c)^{\alpha-2}$\\
                       &$g(\alpha, H)=-\frac{3-\alpha}{4-\alpha}(Hr_c)^{\alpha-2}$\\
\hline
Clausius       &$f(\alpha, H)=-\frac{\alpha}{2} (Hr_c)^{\alpha-2}$\\
                       &$g(\alpha, H)=-(Hr_c)^{\alpha-2}$\\
\hline
\end{tabular}
\label{table1}
\end{table}

In the case of the logarithmic correction the two approaches
differ by a logarithmic term that comes from the $N dT$
contribution. Moreover, contrary to Ref. \cite{zhang10}, we
believe it should not be neglected because $\ln{(A_{AH}/l^2_p)}$
is larger than zero when the area of the apparent horizon
is of order of the Planck area.\\
On its part, power-law entropy correction gives the same
correction to the Friedmann equations, the only difference lying
in the value of the coefficients. The coefficient  $r_c$, there,
is related to the dimensionless parameter $\alpha$ by
$K_\alpha=\alpha (4\pi)^{\alpha/2-1}/(4-\alpha)r_c^{2-\alpha}$.

As can be noted, the $\alpha$ parameter directly comes from quantum corrections to
the entropy and it consequently affects cosmological scenarios. Its value depends
on the details of the quantum calculations, and for the time being there is not
agreement on it. The following analysis determines in which intervals this
parameter results compatible with the GSL.
\section{The Generalized second law of Thermodynamics}\label{Generalized second law of Thermodynamics}
Equipped with the entropy expressions (\ref{log}) and
(\ref{power}), we set out to study whether the GSL is satisfied
when the modified Friedmann equations (\ref{H}) and (\ref{Hdot})
are employed.

Since the entropy depends on the area of the apparent horizon,
$A_{AH}\propto H^{-2}$, it varies as
$$
\dot{S}_H\propto F(H) \dot{H}.
$$
Using eq.(\ref{Hdot}), it can be cast in terms of the Hubble parameter and the
energy density and pressure of the fluid that fills the universe
\begin{equation}\label{entropyrate}
\dot{S}_H=\mathcal{K}\frac{\mathcal{F}(H)}{H^3}\left(\rho+\frac{P}{c^2}\right),
\end{equation}
where
$$
\mathcal{K}=\frac{8\pi^2 c^5 k_B}{\hbar}
$$
and $\mathcal{F}(H)$ depends on the entropy corrections and the thermodynamic relation
used to derive Friedmann equations, namely
\begin{equation}\label{mathFlog}
\mathcal{F}(H)=\frac{1+\frac{\alpha l_p^2 H^2}{\pi c^2}}{1+f(\alpha, H^2)},
\end{equation}
for logarithmic entropy corrections, and
\begin{equation}\label{mathFpower}
\mathcal{F}(H)=\frac{1+\alpha(r_c H)^{\alpha-2}/2}{1+f(\alpha, H)},
\end{equation}
for power-law corrections.\\For the sake of clarity in what follows we split
the analysis for the two classes of entropy corrections, but in this section
we will only consider perfect fluids assuming that the dominant energy condition (DEC)
holds true  (i.e., $\rho+P/c^2>0$) all along the expansion. Then, in view of
eq.(\ref{entropyrate}) the GSL is satisfied provided $\mathcal{F}(H)$ is
non-negative which occurs only for some values of the parameter $\alpha$. In all
the cases the results of general relativity \cite{davies87} are recovered,
as it should, by setting $\alpha=0$.
\subsection{Logarithmic entropy corrections}
\begin{figure}[htb]
\centering
\includegraphics[width=8cm]{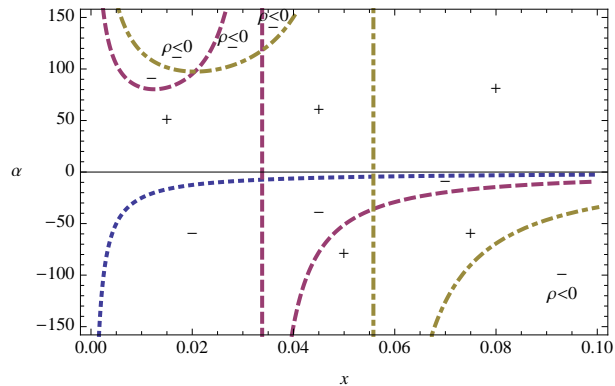}
\caption{Plot of the sign of $\mathcal{F}$ depending on $\alpha$
along the Universe expansion, in the case of the logarithmic
correction and the equipartition principle. Here, $x =
l^2_p/A_{AH}$. The plot focuses on the range $0<x<0.1$ but in the
remaining region the curves behave monotonically. Bear in mind
that the smaller $x$, the older the universe is.}
\label{fig:inequality1}
\end{figure}

\begin{figure}[htb]
\centering
\includegraphics[width=8cm]{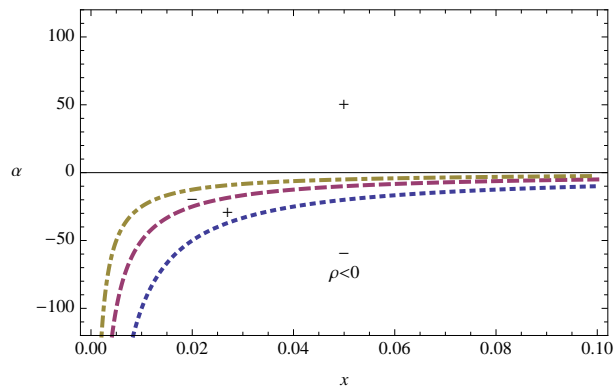}
\caption{Same as Fig. \ref{fig:inequality1}  but with the
equipartition principle replaced by Clausisus relation}.
\label{fig: inequality2}
\end{figure}

For convenience, we introduce the dimensionless variable
$x=l^2_p/A_{AH}$ so that  $x\simeq1$ at the quantum regime and,
provided $\dot{H}<0$, it decreases as time goes on. In terms of
this new variable and using the equipartition theorem,
eq.(\ref{equipartition}), we have that
\begin{equation}\label{Flog1}
\mathcal{F}(\alpha,x)=\frac{1+4\alpha x}{1+2\alpha x\left(1+\frac{1}{2}\ln{4 x}\right)}.
\end{equation}
We require $\mathcal{F}(\alpha,x)$ to be non-negative, if the GSL is to hold,
as well as $\rho\geq0$, not to deal with ghosts. From Friedmann's equation (\ref{H})
with $g(\alpha, H)$ given for the logarithmic correction of entropy, it
translates into the condition
\begin{equation}
\frac{\rho}{H^2}\propto1+g(\alpha, x)=1+\frac{3}{4}\alpha x
\left(1+\frac{2}{3}\ln {4x} \right)\geq 0 \, .
\end{equation}
Fig.\ref{fig:inequality1} depicts the regions in the plane
$(x,\alpha)$ where $\mathcal{F} > 0$ is fulfilled and those in
which $\mathcal{F} < 0$. We believe that the allowed values for
$\alpha$ are just those such that the GSL holds throughout the
expansion of the Universe. The upper bound on $\alpha$ is given by
the local minimum of the dashed curve, that is $\alpha=4 e^3$, and
the lower bound by the intersection of the dotted curve with the
line $x=1$ (i.e., when horizon area equals Planck's area), that is
$\alpha=-1/4$ (not shown).
Positive values of $\alpha$ seem to be largely favored, which is also consistent
with some quantum calculations of the entropy corrections \cite{hod04}.\\
By using Clausius relation (\ref{clausius}), instead, and adopting
the above defined variable $x$ we get
\begin{eqnarray}\label{Flog2}
\mathcal{F}(x,\alpha)&=&\frac{1+4\alpha x}{1+2\alpha x},  \\
\frac{\rho}{H^2}&\propto&1+g(\alpha, x)=1+\alpha x.
\end{eqnarray}
As seen in Fig.\ref{fig: inequality2}, the GSL
is satisfied, $(\mathcal{F}(\alpha,x)\geq0)$, for $\alpha>-1/(4x)$.\\
\subsection{Power-law entropy correction}
\begin{figure}[htb]
\centering
\includegraphics[width=7cm]{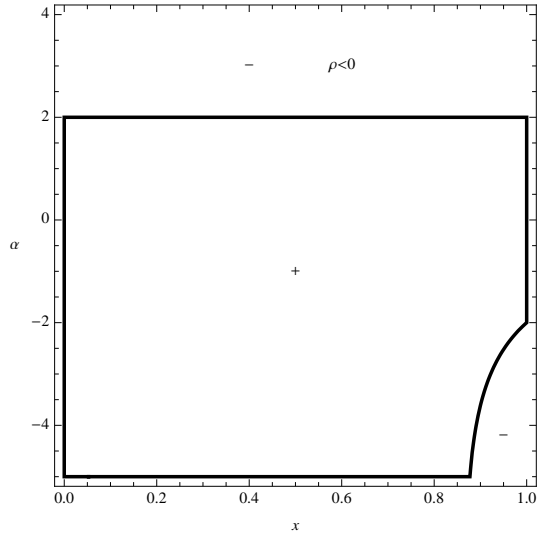}
\caption{Sign of $\mathcal{F}$ depending on $\alpha$ along the
Universe expansion in the case of power-law correction and
Clausius relation. Here $x=H_0/ H$. }. \label{fig:inequality3}
\end{figure}

\begin{figure}[htb]
\centering
\includegraphics[width=7cm]{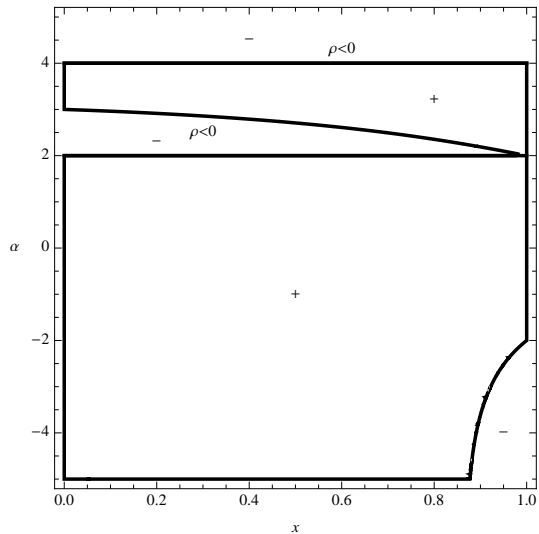}
\caption{Same as Fig. \ref{fig:inequality3} but with Clausius
relation replaced by the equipartition principle.}
\label{fig:inequality4}
\end{figure}

Before applying the GSL let us look at the Friedmann equations for
power-law entropy correction. Inspection of eq.(\ref{power}) shows
that the values of $\alpha= 0$ and $\alpha=2$ are special, in the
sense that for $\alpha=0$ there are no entropy corrections and the
equations reduce to the corresponding general relativity
expressions with a cosmological constant. Likewise $\alpha=2$
represents just a renormalisation of Newton constant, $G$. Bearing
this in mind, we start the analysis by introducing the
dimensionless variable $x=(r_c H)^{-1}$, and identifying the
crossover scale $r_c$ with $H_0^{-1}$, as in \cite{dvali03}. Thus
$x$ tends to zero in the far past and its today value is
$x_0=1$ (provided, again, that $\dot{H}<0$).\\
By virtue of Clausius' relation (\ref{clausius}), the positive
energy condition
$$
1+g(\alpha,x)=1-x^{2-\alpha}\geq0
$$
implies
$$
\alpha\leq2.
$$
Likewise,  the function $\mathcal{F}$ in eq.(\ref{entropyrate}) adopts the form
\begin{equation}\label{F1power}
\mathcal{F}=\frac{1+\alpha x^{2-\alpha}/2}{1-\alpha x^{2-\alpha}/2},
\end{equation}
and it is non-negative in the  ranges
$$
\left\{
\begin{array}{ccc}
\alpha<-2&\text{and}&0<x<\left(-\frac{\alpha}{2}\right)^{(2-\alpha)} \, ,\\
-2\leq\alpha<2&\text{and}&0<x<1 \, ,\\
\end{array}
\right.
$$
as seen in Fig.\ref{fig:inequality3}. The latter shows that the GSL remains
valid throughout the expansion so long as $-2\leq\alpha<2$.
When use is made of the equipartition relation one is led to the
same functional modification of the Friedmann equations, $\propto
H^{\alpha}$, but with different coefficients, which depend on
$\alpha$. In this case, $\rho\geq0$ implies
$$
1+g(\alpha,x)=1-\frac{3-\alpha}{4-\alpha}x^{2-\alpha}\geq0
$$
and
\begin{equation}\label{F2power}
\mathcal{F}=\frac{1+\alpha x^{2-\alpha}/2}{1-\frac{3-\alpha}{4-\alpha}\alpha x^{2-\alpha}}.
\end{equation}
Now, $\mathcal{F}$ is non-negative in the following ranges
$$
\left\{
\begin{array}{ccc}
\alpha<-2&\text{and}&0<x<\left(-\frac{\alpha}{2}\right)^{2-\alpha}\, ,\\
-2\leq\alpha<2&\text{and}&0<x<1,\\
2<\alpha<3&\text{and}&\left(\frac{\alpha(\alpha-3)}{\alpha-4}\right)^{2-\alpha}<x<1 \, ,\\
3\leq\alpha<4&\text{and}&0<x<1 \, .\\
\end{array}
\right.
$$
As can be easily learnt from Fig.\ref{fig:inequality4}, this approach allows a wider range for
the parameter $\alpha$ which can now lie in the range $3\leq\alpha<4$, as well.
\section{GSL with phantom fluid or particle production}\label{GSL with phantom fluid or particle production}
This section investigates whether by relaxing  the dominant energy
condition (DEC) or by allowing particle production, the range of
values of $\alpha$ compatible with the GSL gets wider. The above
analysis shows that this may be the case but there are regions
where the GSL is still violated so long as ghost fields are
excluded. In Figs. \ref{fig:inequality1}--\ref{fig:inequality4}
these regions are explicitly marked.
\subsection{Phantom fluid}
We begin with the logarithmic corrections to the horizon entropy.
First we note that an equivalent rewriting of the entropy rate in
eq.(\ref{entropyrate}), for both eq.(\ref{Flog1}) and
eq.(\ref{Flog2}), is
$$
\dot{S}\propto (1+4\alpha x) \frac{\dot{H}}{H^3}.
$$
Then, for non-superaccelerated expansion $(\dot{H}<0)$, the GSL
will be fulfilled if $1+4\alpha x <0$. This means that we  the
line $\alpha=-1/4$
should not be crossed if a meaningful description of the expansion is required. \\
Nevertheless, we can focus on the positive range of the $\alpha$
parameter in case of equipartition of energy that, as inspection
of Fig.\ref{fig:inequality1} reveals, can be enlarged from the
minimum of the dashed curve, $\alpha=-(2x+x\ln{4x})^{-1}$, to the
minimum of the dot-dashed curve, $\alpha=(3x+2x\ln{4x})^{-1}$, by
allowing the fluid to be phantom during the evolution. These
positive values of $\alpha$ may be seem too big, but they are
not inconsistent with quantum calculations \cite{hod04}.\\
Let us note that, at present, $x_0=l^2_p H^2_0/4\pi c^2\sim 10^{-98}$.
This means that the phantom phase lies far in the past. Could it give
rise to a "reasonable" inflation? The first thing to check is whether
there are accelerated phases in this modified theory of gravity (because
of quantum corrections proportional to $\alpha$). Introducing the parameter
$\gamma$ appearing in the equation of state, $p=(\gamma-1)\rho$, in the
phantom region, when $\gamma<0$ and $\mathcal{F}<0$, positive values
of $\alpha$ are compatible with acceleration $\ddot{a}>0$ provided
that the inequality
$$
 \gamma>\frac{2(1+f(\alpha, H))}{3(1+g(\alpha, H))},
$$
is fulfilled. Then, an inflationary period can be obtained for $4
e^3<\alpha< 8 e^{5/2}$; the latter bound corresponds to the
minimum of the dot-dashed curve in Fig.\ref{fig:inequality1}.\\
For this early inflation to be successful in solving the problems
of the standard big-bang model it must yield a sufficient number
of e-folds, $N\sim 60$.  Bearing in mind eqs.(\ref{H}) and
(\ref{Hdot}) we get
\begin{equation}\label{e-folds}
N=\int_{t_i}^{t_f}H dt=-\frac{1}{3}\int_{x_i}^{x_f}\frac{1}{\gamma}\frac{1+f(\alpha,x)}{1+g(\alpha,x)}\frac{dx}{x}
\end{equation}
The question now reduces to finding an appropriate expression for
$\gamma(x)$ such that the field can be phantom for a period that
gives a suitable  number of e-folds, with the initial and final
values of the phantom period being the intersection of the line
of a given $\alpha$ with the appropriate curve in Fig.\ref{fig:inequality1}.\\
In the case of logarithmic entropy correction and equipartition of
energy, Eq.(\ref{e-folds}) can be analytically integrated to give
a constant value of $\gamma$. For example, for $\alpha=90$, one
has,
$$
N(\alpha=90)=60 \rightarrow \gamma\simeq -0.002.
$$
Although this is just a rough estimate it makes clear that, given
an evolution for the equation of state parameter, it suffices that
it slightly crosses the phantom divide-line to get a convenient
amount of inflation.

In the case of power-law entropy corrections, allowing for a
phantom field may enlarge the range of the $\alpha$ parameter
towards negative values, but a deeper analysis of the evolution
equation shows that the requirement that the GSL is fulfilled
forces the Hubble parameter and the dimensionless variable we have
defined, not to cross the curve $x=-(\alpha/2)^{2-\alpha}$ of
Fig.\ref{fig:inequality4}.
In fact, the entropy rate can be written as
$$
\dot{S}=-\left[1+\frac{\alpha}{2} x^{2-\alpha}\right] \dot{x} x,
$$
hence in the region above that curve $\dot{x}$ must be negative, and
in the region below, positive ($\dot{H}<0$ and $\gamma>0$).\\
Actually this depends on the choice of the crossover scale $r_c$
that we have identified as $H_0^{-1}$. Nevertheless, if this also
depends on the parameter $\alpha$, through a factor, the whole
negative region can correspond to a meaningful expansion, with
$\dot{H}<0$ and $\rho+P/c^2>0$; while all non accessible regions
shifting to the positive sector, with a wider range in case the
Clausius relation to be used.\\
Thus, the results strictly depend on the explicit choice of the
crossover scale, once assumed its order of magnitude is about
$H_0^{-1}$; but it seems that the negative region is accessible,
provided a redefinition of the scale is made, while the positive
part remains largely determined by the power-law dependence.
\subsection{Particle production}
As is well known, on a phenomenological level particle production
can be described in terms of an effective bulk viscosity
$\Pi=-3\zeta H$, with $\zeta$ the coefficient of bulk viscosity
\cite{zeldovich70, hu82, barrow88, zimdahl93}. Thus the total pressure is
\begin{equation}\label{viscousp}
P=p+\Pi,
\end{equation}
where $p=(\gamma-1)\rho$, denotes the equilibrium pressure. In  the case
of a isentropic particle production there exists a general relation
between the viscous coefficient and the particle production rate
\cite{calvao92,zimdahl93}. In this effective picture, in which
particles are accessible to a perfect fluid description soon after
their creation, eqs. (\ref{Hdot}) and (\ref{conservation}) stay as
they are only that $P$ is now given by  eq.(\ref{viscousp}). Likewise,
the entropy rate acquires a new term entirely due to the increase
in the number of particles:
\begin{equation}\label{fluidrate}
\dot{S}_f=\frac{R^3 \Pi^2}{T_f \zeta},
\end{equation}
where $R^3$ is the $3$-spatial volume enclosed by the horizon,
$R^3=4\pi c^3/(3 H^3)$. As before, the fluid is assumed in
thermal equilibrium with the horizon (see eq.(\ref{temperature})).\\
By using eqs.(\ref{entropyrate}) and (\ref{fluidrate}), the total entropy rate is
\begin{equation}
\dot{S}=\dot{S}_H+\dot{S}_f=\mathcal{K}
\left[\frac{\Pi^2}{3\zeta c^2 H^4}+\frac{\mathcal{F}}{H^3}
\left(\rho+\frac{P}{c^2}\right)\right].
\end{equation}
The first term, with particle creation, can allow some region to be reached,
in the sense that GSL can hold with a non-phantom fluid.
Inspection of Figs.\ref{fig:inequality1} and \ref{fig: inequality2}
shows that in case of logarithmic entropy corrections, the $\alpha$
negative range from $-1/4$ to $-1$ could be accessible. In fact,
in this region the field is non-ghost and it was discarded not to
have a superaccelerating evolution so long as $\dot{S}\geq0$.  Moreover,
$\alpha=-1/2$ often appears in the literature \cite{meissner04, domagala04, ghosh04}

\begin{figure}[htb]
\centering
\includegraphics[width=8cm]{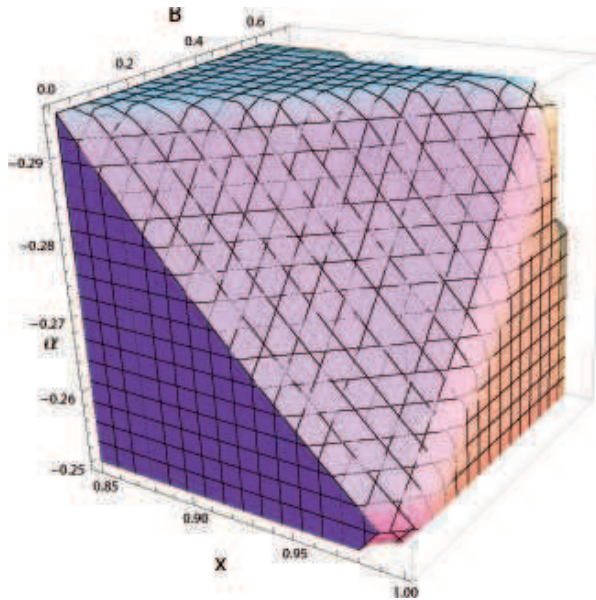}
\caption{Three-dimensional plot of the inequality in
eq.(\ref{entropyratevisc}) in terms of the pair of parameters
(i.e. $\alpha,\,  B$) and the dimensionless  $x$ variable.}
\label{fig: inequalityviscosity}
\end{figure}

Now, because of effective viscosity, a positive term has been included in the entropy rate. \\
So, GSL will hold if
\begin{equation}\label{particleineq}
\mathcal{F}(\alpha,H)\geq\frac{1}{1-\frac{\gamma c^2}{8\pi G \zeta} H (1+g(\alpha,H))}.
\end{equation}
In arriving to this expression we have made use of
$$
\left(\rho+\frac{P}{c^2}\right)=\frac{3\gamma}{8\pi G}H^2(1+g(H))-\frac{3\zeta H}{c^2}.
$$
 In terms of the dimensionless variable $x$, Eq.(\ref{particleineq})  can be written as
\begin{equation}\label{entropyratevisc}
\mathcal{F}(\alpha,x)\geq\frac{B}{B-\gamma\sqrt{x}\left(1+g(\alpha,x)\right)},
\end{equation}
where $B=\zeta \sqrt{16 \pi} G l_p/c^3$.\\
It follows that particle production allows to enlarge the $\alpha$
range from $-1/4\rightarrow-1/2$  in the case of Clausius
relation, and from $-1/4\rightarrow -(2+\ln{4})^{-1}$  in case of
equipartition of energy. This is depicted in the $3$-dimensional
plot of Fig.\ref{fig: inequalityviscosity} that refers to the
latter, and in which a constant value of $\alpha$ can be
maintained all along the expansion, provided a certain amount of
particle production is permitted ($B \neq 0$).

\section{Conclusions}\label{Conclusions}
In this paper we investigated the constraints imposed by the GSL
on modified Friedmann equations that arise from quantum
corrections to the entropy-area relation, eq. (\ref{entropy}). As
is well known, the GSL is a powerful tool to set bounds on
astrophysical  and cosmological models -see e.g.
\cite{paul,macgibbon,chimento,izquierdo}.

Cosmological equations follow either from Jacobson's approach,
that connects gravity to thermodynamics by associating Einstein
equations to Clausius relation (\ref{clausius}), or Padmanabhan's
suggestion that relates gravity (i.e., Einstein equations) to
microscopic degrees of freedom through the principle of
equipartition of energy (\ref{equipartition}). We analyzed two
entropy-area terms, logarithmic (\ref{log}), and power-law
corrections (\ref{power}), the former coming from loop quantum
gravity, the latter from the entanglement of quantum fields.

Both quantum corrections have been widely investigated but, since
they come from very different techniques, one should not be
surprised that total agreement on these corrections is still
missing. In particular, there is a lack of consensus on the value
of the constant parameter $\alpha$. Our work aimed to discriminate
among quantum corrections by requiring, via a classical analysis,
the GSL to
be fulfilled throughout the evolution of the Universe.
This sets constraints on the value of the parameter.

We first investigated the intervals of values of $\alpha$
compatible with the GSL by assuming that the DEC holds true for
the perfect fluid that sources the gravitational field of the FRW
universe. In the case of logarithmic corrections to the horizon
entropy this gives a wide range, in which positive values are
largely favored, with no upper bound in the case of the modified
Friedmann equation derived from Clausius relation. Negative values
of  $\alpha$ are consistent with the GSL  only up to
$\alpha=-1/4$, hence discarding two negative values
that have been suggested in the literature, namely, $\alpha=-1/2$ \cite{ghosh04},
and $\alpha=-3/2$ \cite{kaul00}.\\
In the case of power-law entropy corrections, our analysis favors
positive values of $\alpha$, though they are expected to be
negative. Only a small concordance range appears when modified
Friedmann equations from the equipartition of energy are employed.
Specifically, the interval $3< \alpha < 4$ corresponds to a
power-law correction with an index between $-1$ and 0, that has
been analytically or numerically obtained
\cite{saurya08,jacobson93}.

The second part of our analysis considered the possibility of
enlarging the allowed interval of $\alpha$ in the case of
logarithmic entropy corrections by considering either a phantom
phase for the fluid or particle production modelled as an
effective bulk viscosity. Phantom behavior affects positive values
of $\alpha$ while particle production could enlarge the negative
range and incorporate the value $\alpha=-1/2$ that seems widely
accepted in literature \cite{ghosh04}.

\acknowledgments{NR is funded by the Spanish Ministry of Education
through the "Subprograma Estancias de J\'{o}venes Doctores
Extranjeros, Modalidad B", Ref: SB2009-0056. This research was
partly supported by the Spanish Ministry of Science and Innovation
under Grant FIS2009-13370-C02-01, and the ``Direcci\'{o} de
Recerca de la Generalitat" under Grant 2009SGR-00164.}

\end{document}